\PassOptionsToPackage{table,xcdraw}{xcolor}
\documentclass[runningheads]{llncs}

\usepackage[T1]{fontenc}
\usepackage{tikz}
\usepackage{booktabs}
\usepackage{multirow}
\usepackage{xcolor} 
\usepackage{array}
\usepackage{pifont}
\usepackage{cite}
\usetikzlibrary{arrows.meta, positioning, shapes.geometric}

\tikzstyle{block} = [rectangle, draw, fill=gray!10, text centered, minimum height=1em, text width=5cm, font=\scriptsize]
\tikzstyle{excl}  = [rectangle, draw, fill=red!10, text centered, minimum height=1em, text width=4cm, font=\scriptsize]
\tikzstyle{arrow} = [->, thick, line width=0.5pt]

\usepackage{graphicx}

\begin{document}

\title{SoK: A Systematic Review of Context- and Behavior-Aware Adaptive Authentication in Mobile Environments}

\titlerunning{Survey of Adaptive Authentication in Mobile Environments}

\author{Vyoma Harshitha Podapati~\orcidID{0009-0001-1353-711X} \and
Divyansh Nigam~\orcidID{0009-0001-4765-9449} \and
Sanchari Das~\orcidID{0000-0003-1299-7867}}

\authorrunning{Podapati et al.}

\institute{George Mason University\\
\email{\{vpodapat,dnigam,sdas35\}@gmu.edu}}

\maketitle             
\begin{abstract}
As mobile computing becomes central to digital interaction, researchers have turned their attention to adaptive authentication for its real-time, context- and behavior-aware verification capabilities. However, many implementations remain fragmented, inconsistently apply intelligent techniques, and fall short of user expectations. In this Systematization of Knowledge (SoK), we analyze $41$ peer-reviewed studies since $2011$ that focus on adaptive authentication in mobile environments. Our analysis spans seven dimensions: privacy and security models, interaction modalities, user behavior, risk perception, implementation challenges, usability needs, and machine learning frameworks. Our findings reveal a strong reliance on machine learning ($64.3\%$), especially for continuous authentication ($61.9\%$) and unauthorized access prevention ($54.8\%$). AI-driven approaches such as anomaly detection ($57.1\%$) and spatio-temporal analysis ($52.4\%$) increasingly shape the interaction landscape, alongside growing use of sensor-based and location-aware models.

\end{abstract}

\section{Introduction}
As mobile devices become the primary access point to digital services, growing threats such as credential theft and unauthorized access demand authentication methods that go beyond static credentials~\cite{das2020risk,das2020mfa,kishnani2023assessing}. Adaptive authentication has emerged as a promising solution, dynamically adjusting security mechanisms based on contextual signals such as user behavior, device state, and environmental risk~\cite{arias2019survey,bakar2013adaptive,liu2016privacy}. However, current implementations remain fragmented, often relying on inconsistent design principles and failing to address human-centered concerns. As a result, many systems lack responsiveness to user expectations and contextual nuances~\cite{chen2024mraac,rocha2011a2best,ryu2023design}. Technical advancements frequently outpace improvements in transparency, user control, and trust~\cite{hebbes20112}.

To address these gaps, we present a Systematization of Knowledge (SoK) on adaptive authentication in mobile environments, analyzing $41$ peer-reviewed studies published since $2011$. We identify recurring themes such as continuous and passive authentication, behavioral modeling, and risk-aware decision-making~\cite{papaioannou2022toward,rybnicek2014roadmap}, along with emerging approaches like gesture- and image-based techniques tailored to mobile contexts~\cite{caudill2023wyatt,anand975ai}. Our \textbf{contributions} are twofold: (1) a structured synthesis of over a decade of research on adaptive authentication in mobile environments; and (2) a systematic identification of key challenges, including privacy-preserving models, scalability issues, and the development of a seven-dimension analysis framework.

\section{Methodology}
To structure our review of adaptive authentication on mobile platforms, we adopted the study designs used in prior systematization efforts~\cite{majumdar2021sok,das2019evaluating,noah2021exploring,das2019all,shrestha2022exploring,tazi2024sok,zezulak2023sok,tazi2022sok,duzgun2022sok,shrestha2022sok,huang2025systemization,das2022sok,tazi2023sok,kishnani2023blockchain,saka2023safeguarding,surani2022understanding}. Drawing on recurring technical, behavioral, and ethical themes identified during our preliminary analysis, we formulated the following RQs:

\begin{itemize} \item \textbf{RQ1:} How do adaptive authentication systems incorporate contextual and behavioral signals to enhance security and user experience?

\item \textbf{RQ2:} What ML techniques support real-time decision-making in adaptive authentication, and how are they applied across different systems?

\item \textbf{RQ3:} What are the primary technical and human-centered challenges that affect the design, deployment, and adoption of adaptive authentication in mobile settings? 
\end{itemize}

\subsection{Paper Retrieval and Screening}
\textbf{Search Strategy:} We performed a keyword-based search on Google Scholar using combinations of terms such as~\lq\lq adaptive authentication\rq\rq~and ~\lq\lq mobile devices.\rq\rq~Our search covered publications since $2011$ that returned $342$ papers across the fields of cybersecurity, mobile computing, AI, and HCI. We then collected relevant papers from a variety of digital libraries and repositories. Table~\ref{tab:dist} shows the distribution of papers we gathered from each digital library source.

\begin{table}[ht] 
\centering 
\caption{Distribution of Collected Papers by Publisher} 
\begin{tabular}{|l|r|}
\hline
\textbf{Publisher} & \textbf{Number of Papers} \\
\hline
ieeexplore & 57 \\
dl.acm.org & 11 \\
Springer & 28 \\
Elsevier & 17 \\
academia.edu & 10 \\
arxiv.org & 6 \\
search.proquest.com & 16 \\
researchgate.net & 9 \\
mdpi.com & 12 \\
Other & 86 \\
Unknown Source  & 22 \\
\hline
\textbf{Total} & \textbf{274} \\
\hline
\end{tabular}
\label{tab:dist}
\end{table}
\noindent\textbf{Screening Process:} We applied a multi-phase filtering process involving title, abstract, and full-text screening. We removed $68$ papers for being non-English, inaccessible, duplicates, or lacking core metadata, retaining $274$ for further analysis. Abstract screening eliminated studies unrelated to adaptive authentication in mobile contexts such as those focused only on static biometrics or general security, resulting in $52$ papers.
\\
\textbf{Full-Text Review:} We then evaluated each of the $52$ papers in detail, selecting $41$ that aligned with our criteria: peer-reviewed, English-language, and focused on adaptive authentication for mobile platforms. The final corpus included works addressing system architecture, adaptive mechanisms, user behavior modeling, and risk-based control strategies. The first authors of the paper went through the paper collection and screening process.

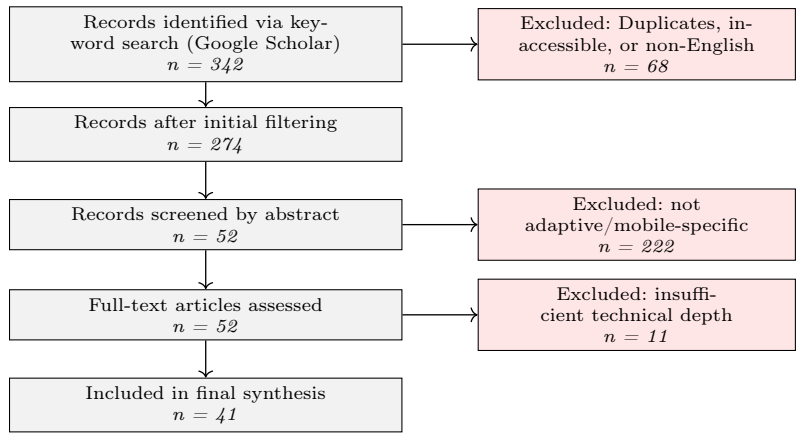
\begin{figure}[ht]
\centering
\begin{tikzpicture}[node distance=1.2cm]

\node (id) [block] {Records identified via keyword search (Google Scholar)\\ \textit{n = 342}};
\node (dup) [excl, right=1cm of id] {Excluded: Duplicates, inaccessible, or non-English\\ \textit{n = 68}};
\node (screened) [block, below of=id] {Records after initial filtering\\ \textit{n = 274}};
\node (abs) [block, below of=screened] {Records screened by abstract\\ \textit{n = 52}};
\node (abs_excl) [excl, right=1cm of abs] {Excluded: not adaptive/mobile-specific\\ \textit{n = 222}};
\node (full) [block, below of=abs] {Full-text articles assessed\\ \textit{n = 52}};
\node (full_excl) [excl, right=1cm of full] {Excluded: insufficient technical depth\\ \textit{n = 11}};
\node (incl) [block, below of=full] {Included in final synthesis\\ \textit{n = 41}};

\draw [arrow] (id) -- (screened);
\draw [arrow] (screened) -- (abs);
\draw [arrow] (abs) -- (full);
\draw [arrow] (full) -- (incl);

\draw [arrow] (id.east) -- (dup.west);
\draw [arrow] (abs.east) -- (abs_excl.west);
\draw [arrow] (full.east) -- (full_excl.west);

\end{tikzpicture}
\caption{Overview of the Paper Retrieval and Filtering Stages Used in the Systematic Review of Adaptive Authentication Studies in Mobile Environments.}
\label{fig:prisma}
\end{figure}

\subsection{Analysis}
To guide our analysis, we developed a comprehensive codebook combining insights from prior surveys with themes that emerged during our review. We organized the codebook around seven core dimensions. Each dimension includes subcodes that capture specific technical strategies and user interaction patterns. The first dimension, \textit{Privacy and Security Models}, focuses on the foundational goals of authentication system design, such as risk-aware mechanisms, continuous behavioral tracking, and device-level access control~\cite{bonazzi2011security,shahzadienhancing,park2025proposal,berbecaru2011lrap}. The second dimension of ~\textit{Interaction Modalities} explores how users engage with authentication systems, including gesture-based inputs~\cite{sejjari2024dynamic,shih2015flick}, sensor-triggered responses~\cite{salomatin2024method,gong2016forgery}, passive background verification~\cite{tanviruzzaman2014your,rahman2014seeing}, and adaptive prompts based on perceived risk~\cite{gupta2018demystifying,seto2014toward}.

Through the third dimension, \textit{User Behavior} we capture when and how authentication is triggered, encompassing continuous background authentication~\cite{hossainpattern,caudill2023wyatt}, event-based triggers~\cite{shah2015multi,anand975ai}, and spatio-temporal strategies responsive to location and time~\cite{berbecaru2011lrap,seto2014toward}. In the fourth dimension of the~\textit{Risk Perception}, we address how users and systems assess cybersecurity threats, covering trust, privacy trade-offs, and concerns over false positives and negatives~\cite{alizadeh2016authentication,shahzadienhancing,tanviruzzaman2014your,park2025proposal}.

In the \textit{Implementation Challenges} we document the barriers to effective deployment, including user resistance to frequent prompts, lack of algorithmic transparency, and hardware or sensor limitations~\cite{agrawal2022study,gong2016forgery}. Via the sixth dimension of~\textit{Usability Needs} we examine how systems accommodate user expectations, focusing on interface design~\cite{gupta2018demystifying,kamarudin2016authentication}, adaptive security behaviors~\cite{anand975ai,rahman2014seeing}, and multimedia-based authentication~\cite{shah2015multi,vazquez2016face}. Finally, in the~\textit{Machine Learning Frameworks} we cover the algorithms supporting adaptive authentication, including behavioral modeling~\cite{caudill2023wyatt,anand975ai}, real-time risk scoring~\cite{shahzadienhancing,hossainpattern}, anomaly detection~\cite{salomatin2024method,ahmadnavigating}, and hybrid AI architectures~\cite{anand975ai,shahzadienhancing}. From a total of $175$ subcategories, we report the $35$ most impactful based on their contribution to key thematic dimensions (Table~\ref{tab:overall}). We collaboratively refined the codebook and piloted it on ten papers using axial and thematic coding, achieving strong inter-coder reliability (Cohen’s Kappa = $0.85$) before applying it to the full dataset.

\section{Results and Discussion}

\begin{table}[!ht]
\centering
\caption{Distribution of Subcategories Of the Seven Focus Areas Based on their Impact Vector}
\renewcommand{\arraystretch}{1.05}
\rowcolors{2}{gray!10}{white}
\begin{tabular}{>{\raggedright\arraybackslash}p{4cm} >{\raggedright\arraybackslash}p{6cm} >{\raggedleft\arraybackslash}p{2cm}}
\toprule
\textbf{Category} & \textbf{Sub-Code} & \textbf{Percentage (\%)} \\
\midrule

\textbf{} & Machine Learning-Based Authentication & 64.29 \\
\textbf{} & Continuous Authentication & 61.90 \\
\textbf{Privacy and Security Models} & Unauthorized Access Prevention & 54.76 \\
\textbf{} & Multi-Factor Authentication (MFA) & 45.24 \\
\textbf{} & Context-Aware Authentication & 42.86 \\

\textbf{} & AI-Based Anomaly Detection & 57.14 \\
\textbf{} & Spatio-Temporal Analysis & 52.38 \\
\textbf{Interaction Modalities} & Background Validation & 50.00 \\
\textbf{} & Adaptive Security Models & 40.48 \\
\textbf{} & Sensor-Based Authentication & 40.48 \\

\textbf{} & Location-Based Adjustments & 47.62 \\
\textbf{} & Risk-Adaptive Authentication Frequency & 45.24 \\
\textbf{Usage Behavior} & Movement Validation & 40.48 \\
\textbf{} & Real-Time Interaction-Based Security & 40.48 \\
\textbf{} & Seamless Reauthentication & 33.33 \\

\textbf{} & Privacy-Preserving Models & 66.67 \\
\textbf{} & Fatigue Mitigation & 52.38 \\
\textbf{Implementation Challenges} & Authentication Complexity Scaling & 50.00 \\
\textbf{} & Frictionless Authentication & 40.48 \\
\textbf{} & Regulatory Compliance & 28.57 \\

\textbf{} & Context-Aware Prompts & 54.76 \\
\textbf{} & Passive Biometric Integration & 52.38 \\
\textbf{Usability Needs} & Non-Intrusive Authentication Mechanisms & 47.62 \\
\textbf{} & Movement Recognition & 28.57 \\
\textbf{} & Hybrid Biometric \& Behavioral Authentication & 21.43 \\

\textbf{} & Privacy-Preserving Authentication Solutions & 76.19 \\
\textbf{} & Behavioral Tracking Concerns & 50.00 \\
\textbf{Risk Perception} & False Positives in ML Authentication & 38.10 \\
\textbf{} & Ethical AI Practices & 23.81 \\
\textbf{} & Surveillance Concerns & 19.05 \\

\textbf{} & Access Control Decisions & 45.24 \\
\textbf{} & Behavioral Classification & 35.71 \\
\textbf{Machine Learning Frameworks} & Risk Score Computation & 33.33 \\
\textbf{} & Context-Aware Frameworks & 30.95 \\
\textbf{} & Behavioral Analysis & 30.95 \\
\bottomrule
\end{tabular}
\label{tab:overall}
\end{table}

\subsection{Behavioral Security Architectures and Risk-Based Control}
Our analysis reveals a significant shift from traditional rule-based security towards behavior-driven and risk-adaptive models. One of the  most common feature among the $41$ studies was \textit{Unauthorized Access Prevention}, cited in 54.8\% of papers. These implementations emphasize active threat detection through monitoring behavioral deviations or unauthorized context switching (e.g., unusual app access or device handoff). Machine Learning-Based Authentication, found in 64.3\% of the papers, highlighting the field’s strong interest in adaptive, environment-aware authentication methods.~\textit{Risk-Based Access Control} is a suggested access-control and authentication mechanism, however it was only applied in 40.5\% of papers, where the researchers' didn't use real-time scoring models to elevate access restrictions. 

A notable trend was the adoption of \textit{Context-Aware Authentication} (42.8\%), in which the authentication mechanism adapts based on user location, activity state, or network conditions. Systems that paired this with \textit{Passwordless Authentication} (40.5\%) and \textit{Continuous Authentication} (61.9\%) emphasizing the importance of ongoing identity verification. These behavior-aware security models form the backbone of most modern adaptive authentication architectures. They often operate in tandem with environmental sensing modules, continuously evaluating biometric, kinetic, and spatio-temporal patterns to determine access legitimacy with minimal friction. This progression reflects an increasing emphasis on dynamic, user-centric security frameworks that respond to real-time behavioral and contextual cues. By integrating multiple signals, these architectures aim to balance robust security measures with seamless user experience.

\subsection{Sensor Fusion and Passive User Interaction}
On the interaction layer, adaptive systems rely on multimodal passive input to evaluate user authenticity. \textit{Sensor-Based Authentication} was the one of the cited technique in this category (40.5\%), leveraging IMU sensors (accelerometer, gyroscope, magnetometer) to extract implicit motion signatures associated with legitimate users. \textit{Continuous Biometric Verification} was used in 26.2\% of systems, combining facial analysis, gait recognition, keystroke dynamics, and touch interaction patterns for persistent authentication. Systems often employed local caching and edge-based analysis to ensure low-latency performance while respecting computational limits on mobile devices.

\begin{table*}[hbt!]
\resizebox{\textwidth}{!}{ 
\begin{tabular}{b{6em} b{8em} *{29}{m{0.52cm}}}
\textbf{Category} & \textbf{Name} \rule{0pt}{19.8em} &
\rotatebox[origin=b]{60}{\makebox[0pt][l]{\footnotesize\textbf{Machine Learning-Based Authentication}}} &
\rotatebox[origin=b]{60}{\makebox[0pt][l]{\footnotesize\textbf{Continuous Authentication}}} &
\rotatebox[origin=b]{60}{\makebox[0pt][l]{\footnotesize\textbf{Unauthorized Access Prevention}}} &
\rotatebox[origin=b]{60}{\makebox[0pt][l]{\footnotesize\textbf{Multi-Factor Authentication (MFA)}}} &
\rotatebox[origin=b]{60}{\makebox[0pt][l]{\footnotesize\textbf{AI-Based Anomaly Detection}}} &
\rotatebox[origin=b]{60}{\makebox[0pt][l]{\footnotesize\textbf{Spatio-Temporal Analysis}}} &
\rotatebox[origin=b]{60}{\makebox[0pt][l]{\footnotesize\textbf{Background Validation}}} &
\rotatebox[origin=b]{60}{\makebox[0pt][l]{\footnotesize\textbf{Adaptive Security Models}}} &
\rotatebox[origin=b]{60}{\makebox[0pt][l]{\footnotesize\textbf{Location-Based Adjustments}}} &
\rotatebox[origin=b]{60}{\makebox[0pt][l]{\footnotesize\textbf{Risk-Adaptive Authentication}}} &
\rotatebox[origin=b]{60}{\makebox[0pt][l]{\footnotesize\textbf{Location-Based Authentication Adjustments}}} &
\rotatebox[origin=b]{60}{\makebox[0pt][l]{\footnotesize\textbf{Risk-Adaptive Authentication Frequency}}} &
\rotatebox[origin=b]{60}{\makebox[0pt][l]{\footnotesize\textbf{Movement-Based User Validation}}} &
\rotatebox[origin=b]{60}{\makebox[0pt][l]{\footnotesize\textbf{Real-Time Interaction-Based Security}}} &
\rotatebox[origin=b]{60}{\makebox[0pt][l]{\footnotesize\textbf{Privacy-Preserving Authentication Models}}} &
\rotatebox[origin=b]{60}{\makebox[0pt][l]{\footnotesize\textbf{Authentication Fatigue Mitigation}}} &
\rotatebox[origin=b]{60}{\makebox[0pt][l]{\footnotesize\textbf{Continuous, Non-Intrusive Authentication}}} &
\rotatebox[origin=b]{60}{\makebox[0pt][l]{\footnotesize\textbf{Gesture \& Proximity-Based Authentication}}} &
\rotatebox[origin=b]{60}{\makebox[0pt][l]{\footnotesize\textbf{Privacy-Preserving Authentication Solutions}}} &
\rotatebox[origin=b]{60}{\makebox[0pt][l]{\footnotesize\textbf{User Concerns Over Behavioral Tracking}}} &
\rotatebox[origin=b]{60}{\makebox[0pt][l]{\footnotesize\textbf{Resilient to Physical Observation}}} &
\rotatebox[origin=b]{60}{\makebox[0pt][l]{\footnotesize\textbf{False Positives in Authentication}}} &
\rotatebox[origin=b]{60}{\makebox[0pt][l]{\footnotesize\textbf{Ethical AI Practices}}} &
\rotatebox[origin=b]{60}{\makebox[0pt][l]{\footnotesize\textbf{Permission Mismanagement Risks}}} &
\rotatebox[origin=b]{60}{\makebox[0pt][l]{\footnotesize\textbf{Behavioral Classification}}} &
\rotatebox[origin=b]{60}{\makebox[0pt][l]{\footnotesize\textbf{Real-Time Risk Score Computation}}} &
\rotatebox[origin=b]{60}{\makebox[0pt][l]{\footnotesize\textbf{Context-Aware Authentication Frameworks}}} \\
\midrule

\multirow{1}{*}{\shortstack{\\Impln.\\Challenges}} & ML-Based~\cite{arias2019survey} &
\ding{108} & \ding{109} & \ding{108} & \ding{108} & \ding{108} & \ding{108} & \ding{109} & \ding{109} & \ding{108} & \ding{109} & - & \ding{109} & \ding{109} & \ding{108} & \ding{108} & \ding{109} & \ding{108} & - & \ding{109} & \ding{108} & \ding{109} & \ding{109} & \ding{108} & \ding{108} & \ding{109} & \ding{109} & \ding{108} \\

& Review~\cite{bakar2013adaptive} &
\ding{108} & \ding{109} & \ding{108} & \ding{108} & \ding{109} &
\ding{108} & \ding{109} & \ding{108} & \ding{108} &
\ding{108} & \ding{109} & - & \ding{108} & \ding{109} &
\ding{108} & \ding{109} & \ding{108} & \ding{109} & \ding{108} &
\ding{108} & \ding{109} & \ding{109} & \ding{109} & \ding{109} &
\ding{108} & \ding{109} & \ding{108} \\

 & Risk-Aware~\cite{papaioannou2021risk} &
\ding{108} & \ding{108} & \ding{109} & \ding{109} & \ding{109} &
\ding{108} & \ding{109} & \ding{108} & \ding{109} &
\ding{108} & \ding{109} & \ding{108} & \ding{109} & \ding{108} &
\ding{109} & \ding{108} & \ding{109} & \ding{108} & \ding{108} &
\ding{109} & \ding{109} & \ding{108} & \ding{108} & \ding{109} &
\ding{108} & \ding{109} & \ding{108} \\

 & ML-Based~\cite{ayyal2023adaptive} &
\ding{108} & \ding{109} & \ding{108} & \ding{108} & \ding{108} & 
\ding{108} & \ding{109} & \ding{108} & \ding{108} & 
\ding{109} & - & \ding{108} & \ding{109} & \ding{108} & 
\ding{108} & \ding{109} & \ding{109} & - & \ding{109} & 
\ding{108} & \ding{109} & \ding{109} & \ding{108} & \ding{109} & 
\ding{108} & \ding{109} & \ding{108} \\

 & IoT-Security~\cite{arab2024towards} &
\ding{109} & \ding{109} & \ding{108} & \ding{109} & \ding{109} &
\ding{108} & \ding{109} & \ding{109} & \ding{109} &
\ding{109} & \ding{108} & \ding{109} & \ding{109} & \ding{108} &
\ding{108} & \ding{108} & \ding{109} & \ding{108} & \ding{109} &
\ding{109} & \ding{109} & \ding{109} & \ding{109} & \ding{108} &
\ding{109} & \ding{109} & \ding{109} \\

 & ML-Based~\cite{liu2024aeaka} &
\ding{108} & \ding{109} & \ding{108} & \ding{109} & \ding{108} & 
\ding{109} & \ding{109} & \ding{108} & \ding{108} & 
\ding{109} & \ding{108} & \ding{109} & \ding{108} & \ding{108} & 
\ding{109} & \ding{109} & \ding{108} & \ding{109} & \ding{109} & 
\ding{109} & \ding{108} & \ding{108} & \ding{108} & \ding{109} & 
\ding{109} & \ding{109} & \ding{108} \\

 & Continuous~\cite{rybnicek2014roadmap} &
\ding{108} & \ding{108} & \ding{109} & \ding{108} & \ding{108} &
\ding{109} & \ding{109} & \ding{109} & \ding{108} &
\ding{109} & \ding{109} & \ding{109} & \ding{109} & \ding{108} &
\ding{109} & \ding{108} & \ding{108} & \ding{109} & \ding{109} &
\ding{108} & \ding{109} & \ding{108} & \ding{108} & \ding{109} &
\ding{109} & \ding{109} & \ding{108} \\

 & Survey~\cite{hasan2025review} &
\ding{108} & \ding{109} & \ding{108} & \ding{108} & \ding{108} & 
\ding{108} & \ding{109} & \ding{109} & \ding{108} & 
\ding{109} & - & \ding{109} & \ding{109} & \ding{108} & 
\ding{108} & \ding{109} & \ding{108} & - & \ding{109} & 
\ding{108} & \ding{109} & \ding{109} & \ding{108} & \ding{108} & 
\ding{109} & \ding{109} & \ding{108} \\

& ML-Based~\cite{shahzadienhancing} &
\ding{108} & \ding{109} & \ding{108} & \ding{108} & \ding{108} &
\ding{108} & \ding{109} & \ding{109} & \ding{108} &
\ding{109} & - & \ding{109} & \ding{109} & \ding{108} &
\ding{108} & \ding{109} & \ding{108} & - & \ding{109} &
\ding{108} & \ding{109} & \ding{109} & \ding{108} & \ding{108} &
\ding{109} & \ding{109} & \ding{108} \\

& Conceptual~\cite{gupta2018demystifying} &
\ding{108} & \ding{109} & \ding{108} & \ding{108} & \ding{108} &
\ding{108} & \ding{109} & \ding{109} & \ding{108} &
\ding{109} & - & \ding{109} & \ding{109} & \ding{108} &
\ding{108} & \ding{109} & \ding{108} & - & \ding{109} &
\ding{108} & \ding{109} & \ding{109} & \ding{108} & \ding{108} &
\ding{109} & \ding{109} & \ding{108} \\

 & Fusion~\cite{rahman2014seeing} &
\ding{108} & \ding{109} & \ding{108} & \ding{108} & \ding{108} &
\ding{108} & \ding{109} & \ding{109} & \ding{108} &
\ding{109} & - & \ding{109} & \ding{109} & \ding{108} &
\ding{108} & \ding{109} & \ding{108} & - & \ding{109} &
\ding{108} & \ding{109} & \ding{109} & \ding{108} & \ding{108} &
\ding{109} & \ding{109} & \ding{108} \\

 & Hardware-Based~\cite{salomatin2024method} &
\ding{108} & \ding{109} & \ding{108} & \ding{108} & \ding{109} &
\ding{108} & \ding{109} & \ding{109} & \ding{108} &
\ding{108} & - & \ding{109} & \ding{109} & \ding{108} &
\ding{108} & \ding{109} & \ding{108} & - & \ding{109} &
\ding{108} & \ding{109} & \ding{109} & \ding{108} & \ding{108} &
\ding{109} & \ding{109} & \ding{108} \\

 & Multi-Factor~\cite{shah2015multi} &
\ding{108} & \ding{109} & \ding{108} & \ding{108} & \ding{109} &
\ding{109} & \ding{108} & \ding{108} & \ding{109} &
\ding{109} & - & \ding{109} & \ding{108} & \ding{108} &
\ding{108} & \ding{109} & \ding{109} & - & \ding{108} &
\ding{109} & \ding{109} & \ding{109} & \ding{108} & \ding{108} &
\ding{108} & \ding{109} & \ding{108} \\

 & Malware~\cite{ahmadnavigating} &
\ding{108} & \ding{109} & \ding{108} & \ding{108} & \ding{108} & 
\ding{108} & \ding{109} & \ding{109} & \ding{108} & 
\ding{109} & - & \ding{109} & \ding{109} & \ding{108} & 
\ding{108} & \ding{109} & \ding{108} & - & \ding{109} & 
\ding{108} & \ding{109} & \ding{109} & \ding{108} & 
\ding{109} & \ding{109} & \ding{108} & \ding{108} \\

 & Factors~\cite{jayabalan2019study} &
\ding{108} & \ding{109} & \ding{108} & \ding{108} & \ding{109} &
\ding{109} & \ding{109} & \ding{109} & \ding{109} &
\ding{108} & - & \ding{109} & \ding{109} & \ding{108} &
\ding{108} & \ding{108} & \ding{109} & \ding{108} & \ding{109} &
\ding{109} & \ding{109} & \ding{109} & \ding{108} & \ding{109} &
\ding{108} & \ding{109} & \ding{109} \\

& Identity-Based~\cite{kamarudin2016authentication} &
\ding{108} & \ding{108} & \ding{109} & \ding{108} & \ding{109} &
\ding{108} & \ding{109} & \ding{109} & \ding{108} &
\ding{109} & - & \ding{108} & \ding{108} & \ding{108} &
\ding{108} & \ding{109} & \ding{109} & - & \ding{109} &
\ding{109} & \ding{109} & \ding{109} & \ding{108} & \ding{108} &
\ding{109} & \ding{109} & \ding{108} \\

 & Location-Based~\cite{berbecaru2011lrap} &
\ding{108} & \ding{109} & \ding{108} & \ding{108} & \ding{108} &
\ding{109} & \ding{108} & \ding{109} & \ding{108} &
\ding{108} & \ding{109} & - & \ding{109} & \ding{109} &
\ding{109} & \ding{109} & \ding{108} & \ding{109} & \ding{109} &
\ding{108} & \ding{108} & \ding{109} & \ding{109} & \ding{108} &
\ding{109} & \ding{108} & \ding{108} \\

 & ZeroTrust~\cite{park2025proposal} &
\ding{109} & \ding{109} & \ding{109} & \ding{108} & \ding{109} &
\ding{108} & \ding{109} & \ding{109} & \ding{108} &
\ding{109} & \ding{109} & \ding{108} & \ding{109} & \ding{109} &
\ding{108} & \ding{109} & \ding{109} & \ding{109} & \ding{108} &
\ding{108} & \ding{109} & \ding{108} & \ding{109} & \ding{109} &
\ding{109} & \ding{109} & \ding{108} \\

 & Biometric~\cite{caudill2023wyatt} &
\ding{108} & \ding{109} & \ding{108} & \ding{109} & \ding{108} & 
\ding{109} & \ding{109} & \ding{108} & \ding{108} & 
\ding{109} & - & \ding{109} & \ding{108} & \ding{109} & 
\ding{108} & \ding{108} & \ding{108} & \ding{109} & 
\ding{108} & \ding{109} & \ding{108} & \ding{109} & 
\ding{109} & \ding{109} & \ding{108} & \ding{109} & \ding{108} \\

\bottomrule

\multirow{1}{*}{\shortstack{\\Privacy \&\\Security\\Models}}& Context-Aware~\cite{liu2016privacy} &
\ding{108} & \ding{109} & \ding{109} & \ding{109} & \ding{108} &
\ding{108} & \ding{109} & \ding{108} & \ding{108} &
\ding{109} & - & \ding{109} & \ding{109} & \ding{108} &
\ding{108} & \ding{109} & \ding{108} & - & \ding{109} &
\ding{108} & \ding{109} & \ding{109} & \ding{108} & \ding{108} &
\ding{109} & \ding{109} & \ding{108} \\

& Privacy~\cite{bonazzi2011security} &
\ding{108} & \ding{109} & \ding{108} & \ding{108} & \ding{109} & \ding{109} & \ding{108} & \ding{108} & \ding{109} & \ding{108} & - & \ding{109} & \ding{108} & \ding{108} & \ding{108} & \ding{109} & \ding{109} & \ding{109} & \ding{109} & \ding{109} & \ding{108} & \ding{109} & \ding{108} & \ding{109} & \ding{109} & \ding{109} & \ding{108} \\

 & Profiling~\cite{baseri2024privacy} &
\ding{108} & \ding{109} & \ding{108} & \ding{109} & \ding{108} & 
\ding{109} & \ding{109} & \ding{108} & \ding{108} & 
\ding{109} & - & \ding{109} & \ding{108} & \ding{109} & 
\ding{108} & \ding{109} & \ding{109} & - & \ding{109} & 
\ding{109} & \ding{108} & \ding{108} & \ding{109} & 
\ding{108} & \ding{109} & \ding{109} & \ding{108} \\

\bottomrule

\multirow{1}{*}{\shortstack{\\Risk\\Perception}} & Risk-Aware~\cite{chen2024mraac} &
\ding{109} & \ding{109} & \ding{109} & \ding{109} & \ding{108} & 
\ding{109} & \ding{109} & \ding{109} & \ding{108} & 
\ding{108} & \ding{108} & \ding{109} & \ding{108} & \ding{109} & 
\ding{109} & \ding{109} & \ding{109} & \ding{109} & \ding{109} & 
\ding{109} & \ding{109} & \ding{108} & \ding{108} & \ding{109} & 
\ding{109} & \ding{109} & \ding{108} \\

& Risk-Based~\cite{papaioannou2022risk} &
\ding{108} & \ding{109} & \ding{108} & \ding{109} & \ding{109} & 
\ding{108} & \ding{108} & \ding{109} & \ding{109} & 
\ding{108} & \ding{109} & \ding{108} & \ding{108} & \ding{109} & 
\ding{109} & \ding{109} & \ding{109} & \ding{108} & \ding{108} & 
\ding{109} & \ding{109} & \ding{108} & \ding{109} & \ding{109} & 
\ding{108} & \ding{109} & \ding{108} \\

& Survey~\cite{papaioannou2023survey} &
\ding{108} & \ding{109} & \ding{108} & \ding{109} & \ding{109} &
\ding{108} & \ding{109} & \ding{108} & \ding{108} &
\ding{109} & \ding{108} & \ding{109} & \ding{108} & \ding{108} &
\ding{109} & \ding{108} & \ding{109} & \ding{108} & \ding{109} &
\ding{109} & \ding{108} & \ding{109} & \ding{108} & \ding{109} &
\ding{109} & \ding{109} & \ding{108} \\

\bottomrule

\multirow{1}{*}{\shortstack{\\Mobile\\Interaction}}& Behavior-Based~\cite{rocha2011a2best} &
\ding{108} & \ding{108} & \ding{108} & \ding{108} & \ding{109} &
\ding{108} & \ding{108} & \ding{109} & \ding{109} &
\ding{108} & \ding{109} & \ding{109} & \ding{109} & \ding{109} &
\ding{108} & \ding{108} & \ding{109} & \ding{109} & \ding{108} &
\ding{108} & \ding{109} & \ding{109} & \ding{108} & \ding{109} &
\ding{109} & \ding{109} & \ding{108} \\

& Biometrics~\cite{agrawal2022study} &
\ding{109} & \ding{108} & \ding{109} & \ding{108} & \ding{108} &
\ding{109} & \ding{109} & \ding{108} & \ding{108} &
\ding{109} & - & \ding{109} & \ding{108} & \ding{109} &
\ding{108} & \ding{108} & \ding{109} & \ding{109} & \ding{108} &
\ding{108} & \ding{108} & \ding{109} & \ding{108} & \ding{109} &
\ding{109} & \ding{108} & \ding{109} \\

 & Biometric~\cite{shih2015flick} &
\ding{109} & \ding{109} & \ding{108} & \ding{108} & \ding{108} & 
\ding{109} & \ding{109} & \ding{108} & \ding{108} & 
\ding{109} & - & \ding{108} & \ding{109} & \ding{108} & 
\ding{109} & \ding{109} & \ding{108} & - & \ding{109} & 
\ding{108} & \ding{108} & \ding{109} & \ding{109} & \ding{108} & 
\ding{109} & \ding{109} & \ding{108} \\

 & Face-Based~\cite{vazquez2016face} &
\ding{108} & \ding{108} & \ding{109} & \ding{108} & \ding{109} & 
\ding{108} & \ding{109} & \ding{108} & \ding{108} & 
\ding{108} & \ding{109} & \ding{108} & \ding{109} & \ding{109} & 
\ding{108} & \ding{108} & \ding{109} & \ding{109} & \ding{108} & 
\ding{109} & \ding{109} & \ding{108} & \ding{108} & \ding{109} & 
\ding{108} & \ding{109} & \ding{108} \\

\bottomrule

\multirow{1}{*}
{\shortstack{\\ML\\Techniques}}& Continuous~\cite{valero2020machine} &
\ding{109} & \ding{108} & \ding{109} & \ding{108} & \ding{109} &
\ding{109} & \ding{109} & \ding{108} & \ding{108} &
\ding{108} & \ding{109} & \ding{108} & \ding{109} & \ding{109} &
\ding{108} & \ding{109} & \ding{108} & \ding{108} & \ding{109} &
\ding{109} & \ding{108} & \ding{109} & \ding{109} & \ding{108} &
\ding{109} & \ding{108} & \ding{109} \\

& ML-Based~\cite{anand975ai} &
\ding{108} & \ding{109} & \ding{108} & \ding{108} & \ding{109} & \ding{108} & \ding{108} & \ding{109} & \ding{108} &
\ding{108} & \ding{109} & \ding{109} & \ding{109} & \ding{109} & \ding{109} & \ding{108} & \ding{108} & \ding{109} &
\ding{109} & \ding{108} & \ding{108} & \ding{109} & \ding{108} & \ding{109} & \ding{108} & \ding{109} & \ding{108} \\

& ML-Based~\cite{hossainpattern} &
\ding{108} & \ding{109} & \ding{108} & \ding{108} & \ding{108} &
\ding{108} & \ding{109} & \ding{109} & \ding{108} &
\ding{109} & - & \ding{109} & \ding{109} & \ding{108} &
\ding{108} & \ding{109} & \ding{108} & - & \ding{109} &
\ding{108} & \ding{109} & \ding{109} & \ding{108} & \ding{108} &
\ding{109} & \ding{109} & \ding{108} \\

& ML-Based~\cite{alizadeh2016authentication} &
\ding{108} & \ding{109} & \ding{108} & \ding{108} & \ding{108} &
\ding{108} & \ding{109} & \ding{109} & \ding{108} &
\ding{109} & - & \ding{109} & \ding{109} & \ding{108} &
\ding{108} & \ding{109} & \ding{108} & - & \ding{109} &
\ding{108} & \ding{109} & \ding{109} & \ding{108} & \ding{108} &
\ding{109} & \ding{109} & \ding{108} \\

\bottomrule

\multirow{1}{*}{\shortstack{\\Usage\\Patterns}} & Transparent~\cite{tanviruzzaman2014your} &
\ding{108} & \ding{108} & \ding{109} & \ding{109} & \ding{108} &
\ding{109} & \ding{109} & \ding{109} & \ding{109} &
\ding{108} & - & \ding{109} & \ding{108} & \ding{109} &
\ding{108} & \ding{109} & \ding{109} & \ding{108} & \ding{109} &
\ding{109} & \ding{108} & \ding{109} & \ding{108} & \ding{108} &
\ding{109} & \ding{109} & \ding{108} \\

 & Habit-Based~\cite{seto2014toward} &
\ding{108} & \ding{109} & \ding{108} & \ding{109} & \ding{108} & 
\ding{108} & \ding{109} & \ding{109} & \ding{108} & 
\ding{109} & - & \ding{109} & \ding{108} & \ding{109} & 
\ding{108} & \ding{109} & \ding{108} & \ding{109} & \ding{109} & 
\ding{109} & \ding{108} & \ding{108} & \ding{109} & \ding{109} & 
\ding{109} & \ding{109} & \ding{108} \\

\bottomrule

\multirow{1}{*}{\shortstack{\\User\\Needs\\Assessment}} & Engineering~\cite{hassan2021engineering} &
\ding{108} & \ding{109} & \ding{108} & \ding{109} & \ding{108} &
\ding{109} & \ding{108} & \ding{109} & \ding{108} &
\ding{108} & \ding{109} & \ding{108} & \ding{109} & \ding{109} &
\ding{109} & \ding{108} & \ding{109} & \ding{108} & \ding{108} &
\ding{109} & \ding{108} & \ding{108} & \ding{109} & \ding{109} &
\ding{108} & \ding{109} & \ding{109} \\

 & Biometric~\cite{ryu2023design} &
\ding{108} & \ding{109} & \ding{108} & \ding{109} & \ding{108} &
\ding{108} & \ding{109} & \ding{108} & \ding{109} &
\ding{108} & \ding{108} & \ding{109} & \ding{109} & \ding{109} &
\ding{109} & \ding{108} & \ding{109} & \ding{108} & \ding{109} &
\ding{109} & \ding{108} & \ding{109} & \ding{109} & \ding{108} &
\ding{109} & \ding{109} & \ding{108} \\

 & Usability~\cite{papaioannou2022toward} &
\ding{109} & \ding{109} & \ding{108} & \ding{108} & \ding{108} & 
\ding{109} & \ding{108} & \ding{109} & \ding{108} & 
\ding{108} & \ding{109} & \ding{108} & \ding{109} & \ding{108} & 
\ding{109} & \ding{108} & \ding{108} & \ding{108} & \ding{109} & 
\ding{109} & \ding{108} & \ding{109} & \ding{108} & \ding{109} & 
\ding{109} & \ding{109} & \ding{108} \\

\bottomrule
\end{tabular}
} 
\noindent
\scriptsize
\textbf{Names:} The mechanism’s title as stated in the paper.\\
\textbf{Evaluation:} \ding{108} = method fulfills criterion; \ding{109} = method quasi-fulfills criterion; -- = method does not fulfill criterion; 
\caption{Impact-Oriented Paper-Wise Distribution of Subcategories Within the Seven Core Focus Areas of Adaptive Authentication.}
\label{tab:results}
\vspace{-8mm}
\end{table*}

\textit{Motion-Based Authentication} (26.2\%) and \textit{Behavioral Deviation Alerts} (14.3\%) played a supplementary role in risk escalation protocols, primarily serving as secondary triggers for additional verification such as challenge prompts or multi-factor authentication when deviations from routine usage patterns were detected. \textit{Seamless Reauthentication Mechanisms} were present in 23.8\% of the systems. These mechanisms continuously updated user state in the background and revoked or adjusted access policies without interrupting workflows. Combined with \textit{Real-Time Interaction-Based Security} (40.5\%), these systems ensured that authentication adapted fluidly to user activity with minimal manual input. The convergence of gesture, movement, and environmental sensing illustrates a shift toward pervasive, real-time verification models in mobile security, designed to operate ubiquitously and invisibly within the mobile ecosystem. This highlights the growing emphasis on continuous, context-aware user validation strategies to enhance both usability and security in adaptive authentication systems.

\subsection{Machine Learning Models for Context and Threat Inference}
Machine learning emerged as the core engine for context classification, behavioral profiling, and threat detection in adaptive authentication.~\textit{Context-Aware Authentication Frameworks} were used in almost all of reviewed systems, enabling devices to synthesize multi-source input including app usage, motion, and temporal behavior into a continuously updating user trust model. This was obvious given the nature of the studies we evaluated.~\textit{Real-Time Risk Score Computation} was used in 33.3\% of studies, typically employing supervised learning classifiers to infer session trustworthiness from time-series features. These scores modulate authentication stringency on a continuous scale, enabling frictionless access in low-risk contexts while invoking challenge mechanisms in high-risk conditions.

\textit{Multi-Factor Risk Analysis} (14.3\%) and~\textit{Fraud Detection in Authentication Requests} (11.9\%) applied ensemble techniques, combining spatial location, sensor activation patterns, biometric traits, and usage trends to detect anomalies indicative of spoofing or unauthorized access attempts.~\textit{Longitudinal Behavioral Data Analysis } featured in 31.0\% of systems, particularly those aiming to establish long-term user baselines. These systems improved detection precision by recognizing gradual changes in usage patterns due to behavioral drift or device sharing. Many papers reported hybrid model architectures: Support Vector Machines (SVMs) were often used for binary classification tasks (e.g., authorized vs. unauthorized), while Random Forests were favored for feature importance modeling. Deep Learning model such as Convolutional Neural Networks (CNNs) for image-based biometrics and Recurrent Neural Networks (RNNs) for time-series modeling were increasingly integrated for multi-modal prediction. This underscores the role of real-time risk scoring as a dynamic mechanism for balancing usability with proactive threat mitigation.

\subsection{Systemic Challenges in Adaptive Implementation}
Adaptive authentication introduces a complex design space requiring trade-offs between performance, accuracy, usability, and trust. One of the cited challenge was~\textit{Adaptive Authentication Complexity Scaling}, observed in 50.0\% of studies. Systems often overcompensated for risk, leading to false positives or over-authentication, which impacted user experience.~\textit{Privacy-Preserving Authentication Models} appeared in 66.7\% of studies, particularly those emphasizing edge-based inference, federated learning, or differentially private model training to minimize exposure of behavioral and biometric data.~\textit{False Positives in Authentication Decisions} were reported in 47.7\% of papers. This issue was most pronounced in systems relying on hard thresholds or static models that failed to accommodate behavioral variability, such as changes in device handling or physical mobility due to injury or environment.

\textit{Regulatory Compliance in Behavioral Tracking} and~\textit{User Consent \& Transparency Measures} were inconsistently implemented across studies, raising ethical questions about silent tracking and unprompted data capture. Papers that addressed these concerns often relied on transparent UI mechanisms or opt-in biometric calibration periods. \textit{Edge Computing for Localized Data Processing} was explored, indicating that its adoption in adaptive authentication remains relatively low. This reflects the continued dominance of cloud-based models, with on-device processing for privacy-preserving authentication yet to achieve widespread implementation. Privacy-preserving models, particularly those using federated learning, offer a promising pathway by enabling collaborative model training across devices without transferring sensitive data to centralized servers.

\subsection{User-Centric Considerations and Ethical Design}
User trust and system transparency emerged as critical factors in determining adoption.~\textit{Continuous, Non-Intrusive Authentication Mechanisms} and~\textit{Passive Biometric Integration} were present in over 50.0\% of the studies. These approaches prioritized background verification to reduce friction and prevent workflow interruption.~\textit{Movement-Based Authentication Recognition} and~\textit{Routine-Based Authentication Learning} were used to tailor the system to individual users, allowing the authentication pipeline to learn and adapt to personal routines while maintaining responsiveness. From a permission standpoint,~\textit{Risk-Based Permission Allocation} and~\textit{Automated Permission Escalation} were employed in 7.0\% of studies, automating access rights based on risk context or app behavior rather than static declarations. However, usability enhancements also introduced privacy tensions.~\textit{User Concerns Over Behavioral Tracking} and~\textit{Surveillance Concerns} were present in over half of the studies. These concerns were especially prominent in systems collecting fine-grained motion or voice data, such as continuous gait recognition or background audio sampling. These findings highlight the delicate balance between delivering seamless authentication and preserving user autonomy, underscoring the importance of ethical design in adaptive systems that learn from behavioral patterns.

\subsection{Perceptions of Risk and Ethical AI in Authentication}
Risk perception plays a pivotal role in the design and acceptance of adaptive authentication.~\textit{Privacy-Preserving Authentication Solutions} were explicitly implemented in 76.2\% of the studies, using cryptographic protocols, on-device inference, and anonymous identity vectors to minimize user profiling. Concerns about explainability and bias were also prominent.~\textit{False Positives in Authentication Decisions} were reported in 38.1\% of the studies, often due to rigid models that failed to adjust thresholds in real time. These errors not only increased user frustration but also eroded system trust.~\textit{Ethical AI Practices} were addressed in only 23.8\% of papers, indicating a gap in the responsible deployment of machine learning in security contexts. Where mentioned, these practices focused on explainable decisions, demographic fairness, and post-deployment auditing.

\textit{User Control Over Permission Revocation} and~\textit{Opt-Out \& Privacy-Safe Alternatives} appeared sporadically, suggesting the need for greater emphasis on transparency and consent in future research. The lack of such controls may hinder the adoption of otherwise technically sound systems, particularly in sensitive sectors like healthcare and border security. Table~\ref{tab:results} provides overview of the papers that focused on the different aspects of our studied elements. The limited focus on ethical AI practices reveals a critical research gap, emphasizing the need for adaptive authentication systems that not only perform accurately but also align with evolving societal expectations around fairness, accountability, and transparency.

\section{Implications}
In this work, we report on adaptive authentication research for mobile platforms and highlights key implications for design, user trust, policy, and future development. Based on our analysis, we offer the following recommendations.

\subsection{Designing Context-Aware Security Frameworks}
The convergence of passive biometric sensing, spatio-temporal data, and machine-learned behavior profiling presents a powerful design paradigm for adaptive authentication. However, the overreliance on static decision thresholds and rigid rule-based triggers (as seen in 47.7\% of systems encountering false positives) underscores the need for adaptive threshold calibration mechanisms that dynamically evolve based on longitudinal usage and contextual cues. Moreover, the frequent adoption of real-time risk-based access control (found in over 40.5\% of systems) illustrates the transition away from one-size-fits-all security models. Developers and system architects should prioritize modular risk engines that integrate seamlessly with mobile OS-level services, offering scalable protection with granular, behavior-conditioned control.

\subsection{Balancing Privacy, Utility, and Transparency}
While over 66.6\% of systems implemented some form of privacy-preserving authentication model, our review highlights a persistent gap in transparency and user agency. Less than 40\% of studies explicitly addressed mechanisms for user-informed consent, opt-out paths, or permission revocation interfaces. This exposes a critical trust bottleneck in the deployment of always-on security systems. Future implementations must address this by incorporating explainable authentication pipelines communicating what data is collected, how risk is assessed, and when access decisions are altered. Integrating user-adjustable sensitivity profiles and visual feedback mechanisms may mitigate behavioral surveillance concerns while improving user acceptance.

\subsection{Operationalizing Adaptive Authentication at Scale}
Despite the promising use of hybrid ML models including CNNs, RNNs, and ensemble methods, practical deployment remains hindered by edge limitations, battery constraints, and inconsistent sensor reliability. Only 50\% of papers addressed on-device inferencing strategies, despite growing user concerns over cloud-based behavioral profiling. This implies a critical need for lightweight, on-device ML architectures, particularly in resource-constrained mobile environments. Federated learning, compressed neural networks, and edge-optimized anomaly detection present viable avenues for future research. In tandem, system developers should engineer fail-safe reauthentication fallbacks to ensure robustness in case of model failure or sensor dropout.

\subsection{Policy and Standardization for Ethical Security AI}
The review identifies inconsistent regulatory compliance across studies, less than 30\% of systems acknowledged frameworks like GDPR, HIPAA, or CCPA, despite their relevance to biometric and behavioral data. As adaptive authentication increasingly relies on sensitive spatio-temporal and motion data, adherence to evolving legal standards becomes essential. To align system design with public interest, researchers and policymakers must co-develop standardized audit frameworks and accountability protocols for AI-based mobile authentication systems. Moreover, only 28.6\% of systems discussed fairness, explainability, or bias mitigation, highlighting an urgent call for algorithmic transparency guidelines and periodic ethical reviews of deployed models, especially in high-risk domains such as border control, finance, and health access systems.

\section{Future Work and Limitations}
We focused our review on English-language, peer-reviewed literature, which may have excluded valuable research in other languages. To broaden the scope, we will incorporate multilingual sources and perspectives from underrepresented regions. We also found that many studies relied on pre-collected datasets or simulations rather than real-world deployments. To address this, we will prioritize in-situ evaluations and contribute to the development of standardized benchmarks. 
\section{Conclusion}
We conducted a systematic review of $41$ peer-reviewed studies on adaptive authentication in mobile environments and identified clear trends in technical design, user interaction, and ethical integration. Our results show that while 64.3\% of systems incorporated machine learning-based authentication, only 33.3\% implemented real-time risk score computation. Continuous authentication appeared in 61.9\% of papers, yet only 23.8\% supported seamless re-authentication mechanisms, revealing gaps in usability and workflow integration. We observed that 52.4\% of studies utilized spatio-temporal analysis, and 57.1\% applied anomaly detection techniques. However, only 30.95\% developed context-aware frameworks capable of adaptive response to user behavior. Privacy-preserving authentication models were reported in 66.7\% of systems, but less than one-third addressed regulatory compliance or user transparency. Additionally, while false positives impacted 38.1\% of implementations, few systems introduced mechanisms for dynamic threshold calibration or behavioral drift adaptation. Our analysis highlights a field advancing in technical sophistication, yet limited by inconsistent support for user consent, explainability, and deployment scalability. 

\section {Acknowledgement}
We gratefully acknowledge the anonymous reviewers for their valuable feedback and insightful suggestions, which have significantly enhanced the clarity and rigor of this systematization of knowledge. We extend our thanks to George Mason University for its institutional support and the resources that facilitated this comprehensive literature review. In addition, we express our gratitude to the researchers whose foundational work on adaptive authentication has been critical to this study; their pioneering contributions continue to shape and advance the evolving landscape of mobile security.

\bibliographystyle{splncs04}
\bibliography{Haisa}
\end{document}